# Context Query Simulation for Smart Carparking Scenarios in the Melbourne CDB

Shakthi Weerasinghe, Arkaday Zaslavsky, Alireza Hassani, Seng W. Loke, Alexey Medvedev, Amin Abken
School of Information Technology
Deakin University
Burwood, Australia
{syweerasinghe, arkady.zaslavsky, ali.hassani, seng.loke, alexey.medvedev, a.abkenar}@deakin.edu.au

*Abstract*-The rapid growth in Internet of Things (IoT) has ushered in the way for better context-awareness enabling more smarter applications. Although for the growth in the number of IoT devices, Context Management Platforms (CMPs) that integrate different domains of IoT to produce context information lacks scalability to cater to a high volume of context queries. Research in scalability and adaptation in CMPs are of significant importance due to this reason. However, there is limited methods to benchmarks and validate research in this area due to the lack of sizable sets of context queries that could simulate real-world situations, scenarios, and scenes. Commercially collected context query logs are not publicly accessible and deploying IoT devices, and context consumers in the real-world at scale is expensive and consumes a significant effort and time. Therefore, there is a need to develop a method to reliably generate and simulate context query loads that resembles real-world scenarios to test CMPs for scale. In this paper, we propose a context query simulator for the context-aware smart car parking scenario in Australia's Melbourne Central Business District. We present the process of generating context queries using multiple real-world datasets and publicly accessible reports, followed by the context query execution process. The context query generator matches the popularity of places with the different profiles of commuters, preferences, and traffic variations to produce a dataset of context query templates containing 898,050 records. The simulator is executable over a seven-day profile which far exceeds the simulation time of any IoT system simulator. The context query generation process is also generic and context query language independent.

*Keywords*-Context Query Simulation, Scalable Systems, Context Management Platforms, Internet of Things

## I. INTRODUCTION

The rapid advancement of the Internet of Things (IoT) has brought about new horizons for development of smarter applications. Context-awareness is an areas of research in this respect, in which the IoT data are fundamental to derive "any information that can be used to characterize the situation of an entity" [1]. These situations, scenarios and scenes that occur in the environment are a result of interactions among different entities – people, places, or any other object which IoT devices and applications sense. With about 75 billion connected smart devices by 2025 [2], and the big IoT data [3] produced as a result, previous work have indicated the immense potential of using context information to make more relevant, informed decisions. Hence, the research in Context Management Platforms (CMPs) is of significant importance as an enabler to massive cyber physical systems.

There are several notable work in CMPs: FIWARE Orion [4], Context Information Management [5], and Context-as-a-Service (CoaaS) [6] platforms. Previous literature about these systems suggests the use of use-case specific data generation processes for testing. For example, FIWARE-Orion is a commercial CMP which could utilize their system logs to re-produce context query loads for research. However, they are proprietary (i.e., own) and are not available publicly. CoaaS could be seen to perform testing using real-world context providers and consumers, e.g., as in [7], with a limited number of entities. Using real world entities for context query and IoT data generation are expensive and cannot be easily scaled. The drawback of this approach is the inability to test CMPs for scale, which is a research gap that we identify to require further investigation. Scalability of CMPs is an important research question given the features of big IoT data [3], and "big" context query loads. Previous work have investigated adaptive context caching [8]–[12], auto-scaling applications [13] for scalability but the area of research is still in its infancy. Evaluation of these work are either theoretical or incorporates IoT data and query loads that are substantially limited in size to be practically similar to real-world settings.

To address this gap, we propose a context query simulator for a smart car parking scenario. We produce a collection of context queries using the context query generation process. Origins of the queries are set in a 12.37 square kilometre area in Melbourne that cover the entire area of the Central Business District (CBD) and adjacent suburbs. The simulation time of the entire context query profile is seven days (i.e., a week).

The novel contributions of this paper are as follows:

1) we introduce a generalizable process to generate context queries for the smart car parking scenario,
2) we produce a large dataset of 898,050 context query templates customized for the Melbourne Central Business District, and,
3) we implement a context query simulator that can generate and execute a sizable collection of context queries over a longer period of time automatically, exhibiting real world patterns and features.

The rest of the paper is structured as follows. Section II briefly discusses related work in context caching. Section III provides an overview of the smart car parking scenario in the Melbourne Central Business District that is simulated using our context query generator. Section IV introduces to the process and implementation of the context query generator and simulation executor. Finally in Section V, we conclude our work with a discussion about the use cases and future work for which the simulator is useful for.

## II. RELATED WORK

There is a significant body of research investigating the scalability of IoT systems [14]–[19]. Researchers have also investigated the use of Information Centric Networks (ICNs) [20]–[23] and Content Delivery Networks (CDNs) [2] in this respect.

TABLE I summarizes the experimental setup of several recent work in IoT data management. Apart from [15], [24], the majority of the work adopt simulations to perform experiments. Authors use simulators such as ccSim simulator in [21], MAT-LAB in [19], and proprietary IoT network simulators developed in C++[20], Python [14], [16]. The primary difference evident in data gathered from simulated experiments versus real-world data is the scale. For instance, Chatterjee et al. [24] defines a one square kilometre area for experimentation, but only tests with 500 requests. Similarly, Wu et al. [15] deploys only 10-15 devices. On the contrary, although for the benefits of scale, reproducibility, and inexpensiveness, authors have resorted to assumptions that cannot be generalized in practical situations, e.g., monotonicity of sensed values [2], [11], [25]. There is a clear compromise between testing for scale and accurate scenarios.

TABLE I. Experimental setups in previous literature. IS – Is Simulated?

| Previous Literature | Experimental Setup | | IS |
|---|---|---|---|
| | IoT Devices | Request Volume | |
| Wu et al. [15] | 10-25 | Unclear | False |
| Rugerri et al. [2] | 10,000,000 | 720-2,916 | True |
| Zameel et al. [19] | 3 | 50-500 | True |
| Zhang et al. | 30 | 18,000 | True |
| Sheng et al. [17] | 30-120 | Unclear | True |
| Nasehzadeh et al. [14] | 40 | 1,000 | True |
| Meddeb et al. [21] | 4000 | 200 | True |
| Zhu et al. [16] | 50 | 500-6,500 | True |
| Fatele et al. [26] | Mostly 100 | 100,000,000 | True |
| Chatterjee et al. [24] | Unspecified | 500 | False |

There are several traffic datasets generated using SUMO such as Luxemburg SUMO Traffic (LuST) [27], and recently the BurST [28] datasets. Apart from the contextual differences of road use in different regions of the world, the main drawback of using tools such as SUMO or any of the derived datasets is the short simulation time. For instance, LuST is produced over a 24hour horizon and BurST only for 1000 seconds. We found no evidence of previous work that simulates for more than 24hours. Yet, similar traffic simulators are only partly useful in scenarios such a smart car parking using context information. Scenarios simulated in a context query simulator need to incorporate other data as well apart from traffic, e.g., context of the users/commuters.

During our literature search, we find that there exist several works in the area of adaptive context caching (ACOCA) – which is one of the strategies to efficiently scale a CMP. Schwefel et al. [11] theoretically propose adaptive caching strategies for CMPs and validates them using a simple manually regenerative simulator configurable using five parameters. Medvedev et al. [9] investigates a cost-efficient algorithm for adaptive context refreshing and performs evaluation using a small query load simulated in JMeter. The queries focus only on a small number of context entities and attributes, tested only over a short period of time, i.e., several minutes. Similar is the work in Weerasinghe et al. [8]. Fanelli et al. [12] propose a context caching mechanism for disaster areas. The authors simulate a disaster area in real-world of 0.1225 square kilometres and execute predefined set of queries at a rate of 4 request per second, only for 10 minutes. The area considered is significantly small in-comparison to an actual disaster area, or a city in our scenario while the time of context query reception is not long enough to credibly test scalability for different situations. According to [29], context queries and IoT data lacks any pattern. But considering traffic variation patterns as indicated in [30], complete randomly generated or pre-defined context query loads using tools such as JMeter do not reflect the real-world scenarios or could be used to configure in such a manner.

The smart car parking use case has previously been used for testing by Hassani et al. [7]. The authors evaluate the Context Services Description Language (CSDL) using live data retrieved from ten car parking areas and a single context consumer. Although the practical setup can produce the most reliable data and situations, there are drawbacks: (a) expensiveness; for instance, the vehicle used by the context consumer in [7] is modified to communicate with the mobile application, (b) difficulty in scaling because of the necessary customizations, infrastructure costs, and registering a sizable number of context services and consumers, and (c) limited number of entities, and volume of context queries, compared to a real-world scenario, i.e., in a large metropolitan city. Accordingly, we identify that there exists a significant gap in simulating a scenario that is sizable enough, i.e., city-wide, to reliably test the scalability of CMPs and their functionalities, with the least cost and effort.

Rather than using live-data of available slots in parking facilities, Ivan et al. [31] predict the availability of parking facilities using traffic congestion information – an indirect method to produce data using indicators. Using this work as motivation, we noted that the volume of the context queries in search of car parks would also follow the variation of the traffic volume. According to Afrin et al. [32], we further identify that the traffic congestions occur due to high density of vehicles, which intuitively translates to crowdedness of a location. Hassani et al. [33] use real-world datasets, i.e., On-street parking, to demonstrate the scalability of CoaaS. Similar data and statistics are made publicly available in Melbourne which present the feasibility to develop reliable datasets and simulators. Therefore, using the real-world traffic and crowdedness data available via different sources, we developed a strategy to simulate a city-wide smart car parking scenario which is of significant importance to testing scalable CMPs.

## III. SMART CARPARKING IN MELBOURNE

In this section, we introduce the smart car parking scenario in the Melbourne Central Business District (MEL CBD). We present an example of a complex context query involving multiple entities that could be simulated using our method.

### A. Motivation for Context Query Simulation

The primary motivation for developing a context query simulator and a data set is the lack of a sizable context query load that depict the patterns and features of a real-world scenario. A significant body of work in CMPs are highly theoretical or illustrated using limited sized test results. These are not adequately sized to investigate real-time adaptive mechanisms and techniques to reliably evaluate the practical feasibility of the system. It is a significant drawback when performing experimental data gathering in research related to CMPs. We also indicated that generating context queries in a real-world setting is expensive and requires significant amount of effort and time to generate a sizable amount of context queries. Yet, given the transient nature of situations and context, it is impossible to re-create situations and scenarios unless they are tampered upon. It is difficult to reliably perform comparative analysis and/or benchmarking in such as setup. For example, consider comparing the cost and performance efficiency of Adaptive Context Caching (ACOCA) mechanism versus a no caching approach [8]. This creates the necessity to develop a context query simulator that conforms to real world indicators of situations, scenarios, and scenes. The simulator should exhibit the following features: (a) cheap, (b) configurable, (c) ability to reproduce a context query load, (d) produce context query loads that exhibit real-world patterns and features, and (e) takes less time and effort to generate. There is a need for a context query simulation that conforms to real-world indicators in the real-world to test the suitability of any research solution in scalable CMPs and derive credible results.

Previous work has considered either, (a) a limited number of context queries e.g., a single context query in , and/or entities, e.g., a single car park in [8], a single garbage collector and a smart bin in [34], (b) context queries generated from a limited number of context consumers, or (c) simulated with a small number of entities. None of these context query loads are in the scale of the real-world or capable of mimicking real-world patterns and features.

Our context query generator is independent of any query language. Other researchers can utilize our data or process for testing purposes using their own proprietary context query languages and platforms well. Further, our context query generation process for the smart car parking scenario can be generalized for any other region in the world provided similar datasets, and statistics are available.

### B. Overview of the Scenario

Melbourne (MEL) CBD is a metropolitan area with a defined area of 1.79 square kilometres and containing a population density of 26,420 people per square kilometre. Although approximately 80% of the population are residents, a sizable 50% are students. The number of daily commuters including for commercial work, tourism, studies, sporting, and other crowd pulling events (e.g., the AFL Finals, Boxing Day Cricket, Australian Open, Melbourne Cup are typically hosted in Melbourne) to and from the MEL CDB is significant which demands for car parking facilities. Melbourne boasts of about 42,000 carparking spaces [35], but not all car parks are useful in a search result for a commuter based on factors such as preference, distance to the destination, opening hours, price offered, accessibility options, and the friendliness to travel between the car park and the destination (which we refer to as *goodForWalking* in this paper). For instance, a parking spot located 500m away from the destination is *friendly* (i.e., *goodForWalking*) if the weather is calm. It is however not so during a thunderstorm, e.g., as in the example in [33]. Still the same parking spot can be *unfriendly* to another commuter on a calm day given a higher UV index, i.e., medically unsuitable for the user. Therefore, there exists a demand-and-supply problem of "suitable" parking spots. Recommending available parking slots based only on the number of available slots in a parking facility is not sufficient in this situation. Context-aware parking recommendation resolves this "suitability" issue by considering the relevant context of the user (whom we refer to as the context consumer), and other entities, i.e., the car park, and traffic conditions, to provide more relevant search results.

Fig.1 illustrates an example context query in search of available parking slots using the Context Definition and Query Language (CDQL) [33]. The query produces search results based on the context of the car park (i.e., availability, suitability of the available slots in terms of the dimensions of the consumer's vehicle), the context of the context consumer (e.g., preference of price, distance), and the situation of the surrounding environment (e.g., *goodForWalking*). Situations are referred to as the *derived* or *high-level* context which in CDQL is defined as situational functions [33] that implement the Context Space Theory [36].

```
prefix schema:http//schema.org
pull (targetCarpark.*)
define
entity targetLocation is from schema:Place
  where targetLocation.name="Melbourne Skydeck",
entity consumerCar is from schema:Vehicle
  where cosumerCar.vin="13UNVER82367G4",
entity taregtWeather is from schema:Thing
  where targetWeather.location="Melbourne,VIC"
entity targetCarpark is from schema:ParkingFacility
  where
    ((distance(targetCarpark.location,
targetLocation.geo, "walking")<{"value":200,
"unit":"m"} and
goodForWalking(targetWeather)>=0.6 or
goodForWalking(targetWeather)>0.9) and
    targetCarpark.maxHeight>consumerCar.height
and
    targetCarpark.isOpen=true and
    targetCarpark.availableSlots>0 and
    targetCarpark.price<={"value":20,
"unit":"aud"} and
    targetCarpark.rating>=3 and
    isAvailable(tagetCarpark.availableSlots,
{"start_time":now(), "end_time":{"2020-07-
12T18:00:00", "unit":"datetime"}})
```

Fig.1. Context-aware available parking slots query defined in CDQL [33].

In this paper, we present our process of generating a data set of context query templates that get executed using the Context Query Executor (CQE) to simulate a load of context queries for the smart car parking scenario during a span of one week. We produce our context query templates inspired by real-world traffic variations, and popularity of locations in the MEL CBD using actual datasets. Therefore, our Context Query Simulator (CQS) executes context queries that follow near real-world patterns, e.g., the context query volume peaks during the rush hours, and features, e.g., the variation of number of context queries specifying a certain destination by day, expected in a stream of context queries executed with in the MEL CDB. Fig.2 illustrates the scenario used to generate the dataset. The red dots represent destinations of context consumers stated in the context queries and the blue squares with white "P"s represent car parking facilities.

In the next section, we describe our approach to generating the dataset of context query templates having features similar to actual commuters, traffic variations and popularity of places in the MEL CBD and executing them randomly, but proportional to the indicators of context query volume. We use the terms context query templates and dataset interchangeably in this paper.

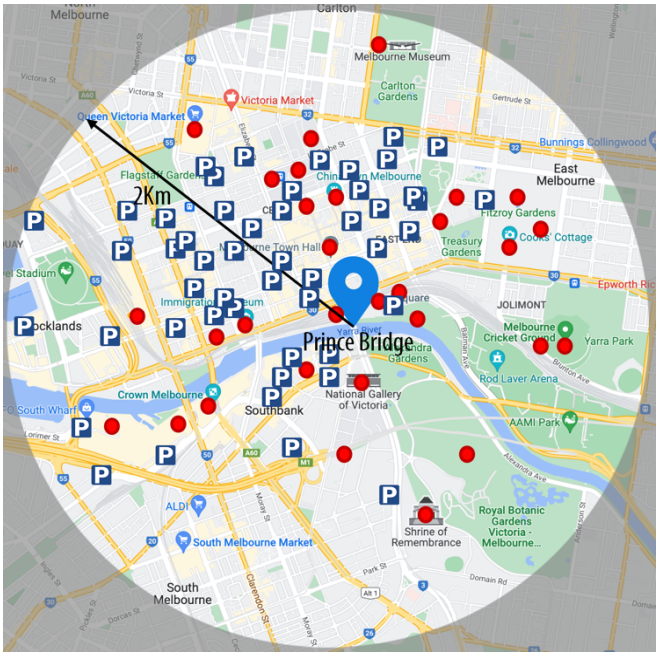

Fig.2. Overview of simulated scenario in MEL CBD.

IV. CONTEXT QUERY GENERATOR AND EXECUTOR

The context query simulator comprises two primary components: (a) Context Query Generator (CQG), and (b) Context Query Executor (CQE). CQG was developed using Python 3.8.2 and CQE was developed using Java 15.0.1. Both the components are containerized in Docker. We used this approach for implementation with the intention of reusing the context query load generated using the CQG over multiple test cases using the CQE. Fig.3 illustrates the high-level architecture of the simulator.

The process of generating the context query load and simulating the context query load for the smart car parking scenario in the MEL CDB follows a three-step process – (a) data gathering, (b) context query generation, and (c) context query execution. We explain them further as follows.

*A. Data Gathering*

There are four datasets that are fundamental to the context query generation process: (a) the popularity of places, (b) traffic volume, (c) car parks, (d) vehicles (i.e., automobile). First, we retrieved the Traffic Volume Dataset[1] in Victoria, Australia and the Automobile Dataset [37]. Next, we scraped the webpages for each search result for car parks using the search key "Melbourne VIC, Australia" in Wilson Parking[2] and Secure Parking[3]. Incomplete and inaccurately collected data about the carparks were removed and cleaned. A total of 61 car parks were collected in this process, located within two kilometres from the Price Bridge (12.37 square kilometres in area which also intersects that adjacent suburbs of the MEL CBD). In order to further improve the quality of data, we enriched each collected record of a car park with the data collected from the Google Places API. We used the text search feature for this purpose, e.g., /textsearch/json?query=Southgate%20Car%20Park.

Finally, we developed a dataset of popularity of places by combining the Google API and the open-source Populartimes web scraper[4]. Interested readers are referred to our repository for further details. Our implementation takes a list of names of places as input and first retrieves the primary details of the location using the text search feature in Google Places API. Then, the reference ID of the place is used by the scraper to derive the popularity variation of the location. As indicated in the example in Fig.4 and Fig.5, this process generates a distribution of average relative popularity by each hour for all seven days of the week. The popularity distribution of places against time is one of the factors affecting the behaviour of the simulated context query load, i.e., the seven-day context query profile, as we will describe in the next sub-section. Our dataset is generated considering thirty popular and/or tourist locations in Melbourne, e.g., Flinders Street Station, Royal Botanical Gardens Victoria, Melbourne Museum, etc.

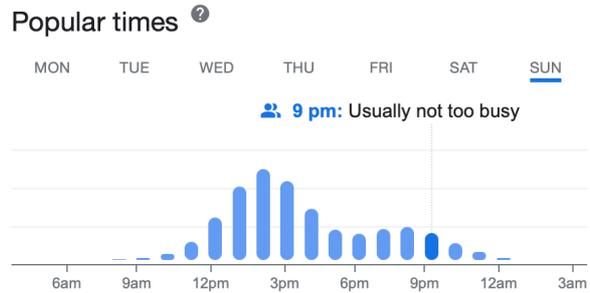

Fig.4. The distribution of average relative popularity of Melbourne Cricket Ground in graphical form in Google.

```
{
  "populartimes":[
    "Monday":[
      9,1,0,0,8,19,31,39,40,37,36,41,50,60,68,7
      2,71,65,55,43,30,19,11],
    "Tuesday":[
      4,0,0,0,8,24,40,49,45,36,32,37,49,63,74,8
      1,82,76,65,51,37,24,13],
    "Wednesday":[
      6,1,0,0,3,11,22,34,44,48,47,41,36,41,57,80,
      97,97,79,54,40,42,41,24],
```

---

[1] Available at: https://discover.data.vic.gov.au/dataset/traffic-volume
[2] Available at: https://www.wilsonparking.com.au
[3] Available at: https://www.secureparking.com.au/en-au/
[4] Available at: https://github.com/m-wrzr/populartimes

```
    "Thursday":[
        5,0,0,0,3,12,23,35,45,49,47,42,39,46,63,85,
        100,98,81,59,47,47,44,27],
    "Friday":[
        9,0,0,0,5,10,19,28,37,41,41,39,38,44,57,73,
        85,85,75,63,56,55,54,48],
    "Saturday":[
        35,21,9,4,4,5,8,15,23,32,42,51,58,63,67,74,
        82,86,80,70,65,66,68,63],
    "Sunday":[
        50,33,18,8,1,5,9,14,20,26,33,42,52,62,70,73
        ,70,62,53,48,46,44,36,22]
    ]
}
```

Fig.5. The distribution of average relative popularity of Flinders Street Station (FST) in JSON form.

### B. Context Query Generation

As indicated in Fig.3, the context query load generation follows a multi-step process. The output of the CQG is a collection of context query templates which we store in the Context Query Template Repository (CQTR). MongoDB is used for this purpose in our work.

First, we perform a statistical analysis of the popularity distribution to derive several aggregate distributions. We define two configuration parameters: (i) multiplier ($\alpha$), and (ii) probability of querying ($P(Q)$). Given the values of the popularity distributions are relative values ($RP$) compared to the highest average size of the crowd gathered on a particular hour, the multiplier is used to convert the values to an absolute number. The probability of querying refers to the proportion of the crowd who have made context queries. We represent it mathematically in (1). Accordingly, the number of context queries (CQ) we generate during hour ($h$) given the destination ($l$) of the context consumer can be given as follows:

$$P(Q) = \frac{Crowd\ executed\ context\ queries}{Total\ Size\ of\ the\ Crowd} \quad (1)$$

$$CQ_{l,h} = RP_{l,h} \times \alpha_{l,h} \times P(Q)_{l,h} \quad (2)$$

For instance, consider that the relative popularity of Flinders Street Station during 12:00-1:00pm is 70 (i.e., 70%). Assuming $\alpha = 5$ and $P(Q) = 0.4$, the total number of context queries generated for simulation during the above time, with the intention of finding car parks near Flinders Street Station is, $70 \times 5 \times 0.4 = 140$. For simplicity, we set $\alpha = 10$ and $P(Q) = 1.0$ in generating our dataset for all locations. If $L$ is the set of all interested places, the total number of context queries ($CQ_{total}$) during the weekly profile can be derived as follows in (3). Accordingly, our dataset contains the 898,050 context query templates generated.

$$CQ_{Total} = \sum_{l=0}^{L} \sum_{h=0}^{23} RP_{l,h} \times \alpha_{l,h} \times P(Q)_{l,h} \quad (3)$$

The statistical analysis process produces three outputs: the distribution of (a) size of the crowd by place, (ii) size of the crowd by day, and (iii) size of the crowd by place and day. Fig.6 illustrates the popularity distribution transformed into the number of context queries by day of the week.

Next, we create the collection of context query templates. These templates are developed incrementally through a series of steps to generate the execution ready set of queries. First, we create a set of empty context query templates, equal to the total number of context queries calculated using (3). Then, our matching algorithm, taking the size of the crowd by day and size of the crowd by place distributions as inputs, assigns the day of the week and the intended destination for the context consumer. The query execution time matcher further attaches the exact time the query needs to be executed. For example, consider that the RP = 80 for Flinders Street Station (FST) during 10:00-11:00am (i.e., $h$=10) and the $CQ_{FST,10}$=80. Accordingly, the expected Poisson request rate ($\lambda$) of context queries containing the intention to search available car parks near FST is 80 per hour. Each of these 80 requests are assigned a random time during this hour up to the second, e.g., 20 seconds past 10:34am. Fig.7 illustrates the number of context queries per minute (i.e., request rate) in our generated dataset for FST during $h$=10 on Monday.

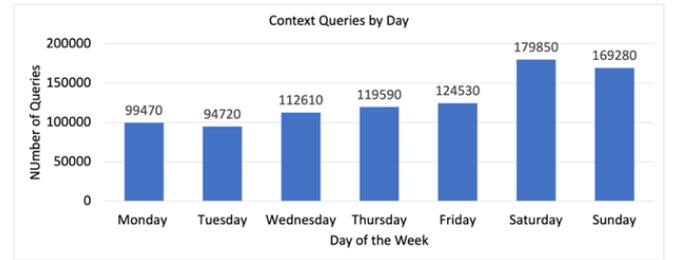

Fig.6. Number of context queries executed by the day of the week.

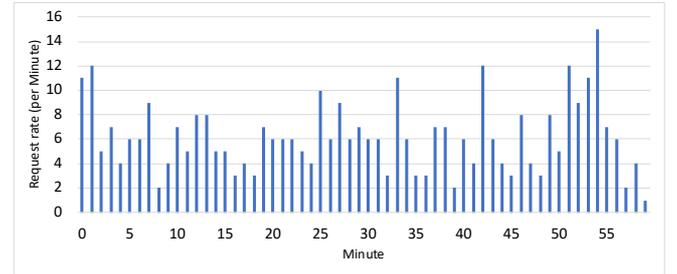

Fig.7. Request rate per hour given the destination defined is FST.

The origin of the context consumer is assigned randomly to each context query template. We implemented a random location generator with a configurable radius. In order to maintain consistency, we set the radius to 2Km when generating our dataset with the centre set to the Prince Bridge. Fig.8 illustrated all the locations from which the context queries are executed by the context consumers.

The next step of matching the context query (having assigned the datetime and destination) with the context consumer is a complicated matching algorithm that match the popularity distribution by time and location with the different context consumer (i.e., commuter) profiles. We identified thirteen commuter profiles by mining the Traffic Volume Dataset. TABLE II summarize the distribution of the context queries among these profiles. In the table, *same_time = True*, refers to commuters who arrive and/or make the context query recurringly in the same hour of the day. The random commuter profile refers to those who do not follow any of the other profiles. For example, a commuter travelling on Monday, Wednesday and Friday at different times are categorized in this profile. We also used the data from the Auditor General of Victoria's report [30], such as the variation of traffic load in Melbourne (Fig.9) and the peak traffic hour statistics to map recurrent commuters (e.g., commuter types - 1, 3, 5, and 7) to queries assigned execution time in the rush hours to simulate the real-world situation.

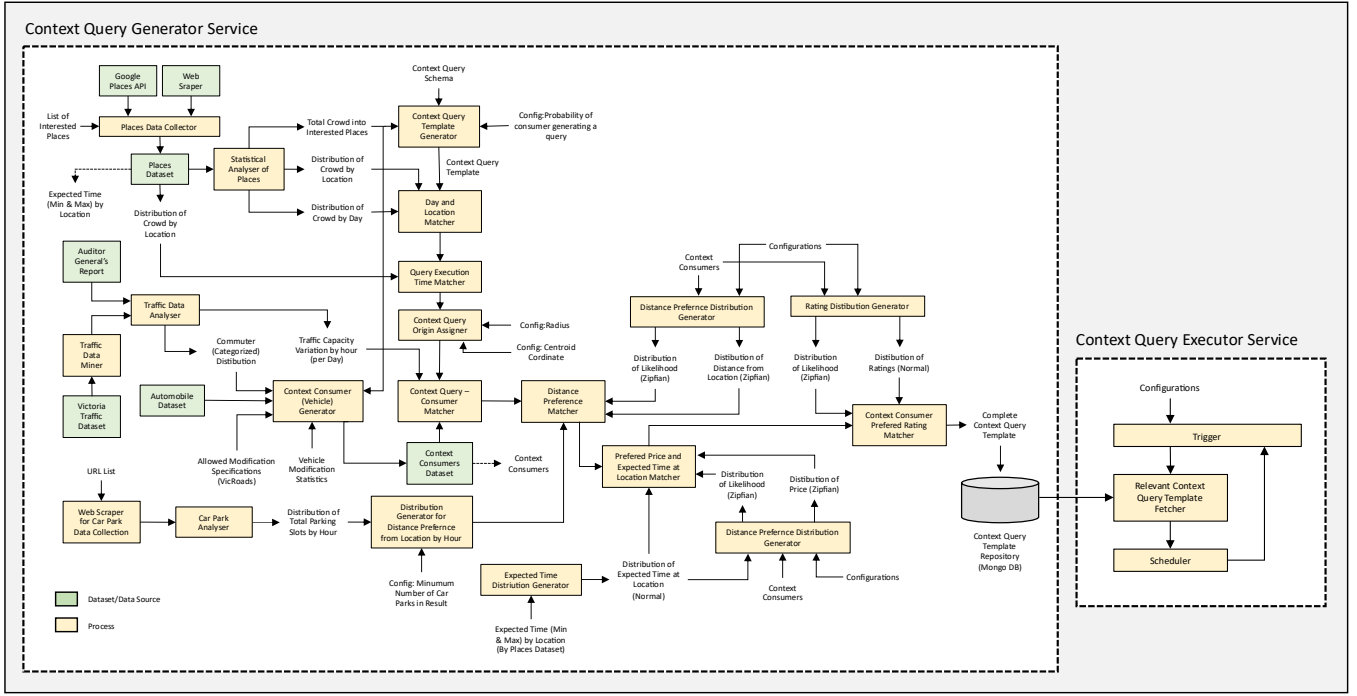

Fig.3. Architecture of the Context Query Simulator for the Smart car parking scenario.

TABLE II. The probability distribution of context queries to each commuter profile. Prob. – Probability, CQs – Context Queries, and NA – Not Applicable

| No | Type | same_location | same_time | Joint Prob. | Number of CQs | Prob. |
|----|------|---------------|-----------|-------------|---------------|-------|
| 1  | Daily Commuters | True | True | 0.022 | 84,000 | 0.0935 |
| 2  |  | True | False | 0.007 | 28,000 | 0.0312 |
| 3  |  | False | True | 0.004 | 14,000 | 0.0156 |
| 4  |  | False | False | 0.004 | 14,000 | 0.0156 |
| 5  | Weekday Commuters | True | True | 0.055 | 150,000 | 0.1670 |
| 6  |  | True | False | 0.018 | 50,000 | 0.0557 |
| 7  |  | False | True | 0.009 | 25,000 | 0.0278 |
| 8  |  | False | False | 0.009 | 25,000 | 0.0278 |
| 9  | Weekend Commuters | True | True | 0.026 | 28,000 | 0.0312 |
| 10 |  | True | False | 0.026 | 28,000 | 0.0312 |
| 11 |  | False | True | 0.003 | 3,500 | 0.0039 |
| 12 |  | False | False | 0.010 | 10,500 | 0.0117 |
| 13 | Random Commuters | NA | NA | 0.805 | 420,050 | 0.4674 |
|    | Total |  |  | 1.000 | 898,050 | 1.000 |

For simplicity, we assume that each commuter is a context consumer given $P(Q) = 1$ in our dataset generation configuration. Based on this assumption, we attach a random vehicle from the Automobile dataset to each commuter. The percentage number of vehicles modified is configurable. We set 0.5 and the specifications of 50% the assigned vehicle are modified in the dataset where each selected parameter for modification are updated according to a normal distribution. The normal distributions are generated taking the range of 0 to maximum allowed legal modification into account. Given the assigned vehicle is selected for modification, each parameter has a probability of 0.5 to be updated. In our dataset generation process, we configured for the modification of height, length, and width of the modified vehicles. Based on this matching process, there are 543,050 different context consumers in our dataset.

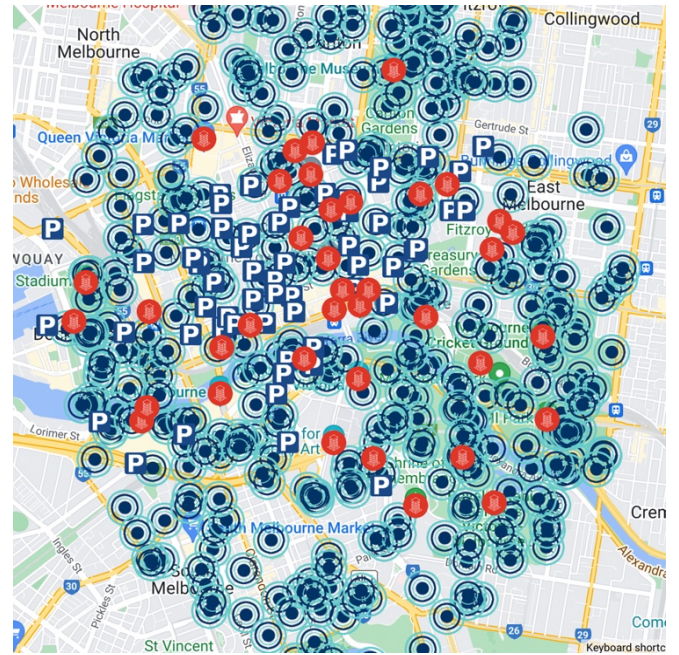

Fig.8. Locations of context query execution during $h$=10 with the destination set to FST.

Given the previous step assigns context consumers the context query template, the rest of the steps attach preference-based context query conditions and parameter values to the context query template. The following preference conditions of the context consumers are attached: (a) maximum distance from destination, (b) price, (c) expected time of parking, and (d) minimum expected average rating. The probability of a context query containing each condition follows a Zipfian distribution individually. We generate them randomly in our algorithm and TABLE III presents the number of context queries attached with each condition and their relative occurrence in our dataset.

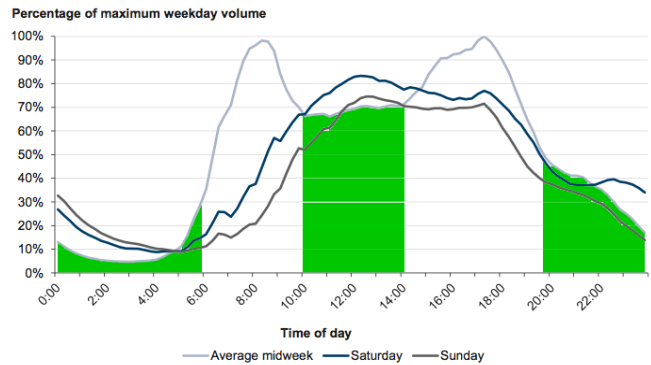

Fig.9. Variation of relative traffic volume in Melbourne [30].

TABLE III. Probability of occurrence of optional conditions in the dataset.

| No | Optional Condition | Probability |
|---|---|---|
| 1 | Preferred minimum average rating | 50.03% |
| 2 | Preferred maximum price to pay | 80.04% |
| 3 | Whether the car park is available to park during the entire expected duration? | 20.03% |

Using the dataset of carpark in MEL CBD, we aggregated the number of available (i.e., open) car parks for each thirty minutes on each day of the week using the car park analyser. The process generates a distribution of average number of available parks by hour. We transform this availability distribution to a distance preference by considering: (a) the configuration that set the minimum number of car parks expected in the context query result, and (b) the distribution of available number of parking slots in each car parks. The variation of available parking slots were simulated using our previous work, i.e., the car park simulator in [8], conforming to a "random" variation. For example, consider there are only 20 car parks open during 01:00-01:30am. For a given destination as the centre, the average radius to contain at least two car parks with available slots in the context query result is 500m. We further aggregated this result into four spans as indicated in TABLE IV using the traffic volume data statistics indicated in [30] for simplicity when generating our dataset. In the table, dominant refers to the majority of the context consumers. For example, during 05:00-10:00am, 80% of the context consumers executing context queries at this time prefer to park at an average maximum distance of 137.5m from the destination.

TABLE IV. Preference of distance to the destination from car park according to time spans. $\mu$ – average distance.

| Time Span | Dominant | Non-Dominant | $\mathbb{E}$[Distance] |
|---|---|---|---|
| 05:00-10:00 | 0.2 $\mu$ =137.5m | 0.05 $\mu$ =1,150m | 85m |
| 10:00-14:00 | 0.125 $\mu$ =189.29m | 0.125 $\mu$ =1,312.5m | 187.72m |
| 14:00-19:00 | 0.2 $\mu$ =137.5m | 0.05 $\mu$ =1,150m | 85m |
| 19:00-01:00 | 0.1 $\mu$ =189.29m | 0.15 $\mu$ =1,312.5m | 215.8m |
| 01:00-05:00 | 0.2 $\mu$ =115m | 0.005 $\mu$ =1,000m | 73m |

Apart from the time of the day, the value of the maximum distance from the destination preferred by the context consumer is impacted by: (a) user's own static preference, e.g., always prefer a distance of 50m, (b) crowdedness of the location, and (c) random choice. We configured each of the factors to decide the value with equal probability. Fig.10 illustrates the joint probability distribution of distance value assignment based on crowdedness of the location.

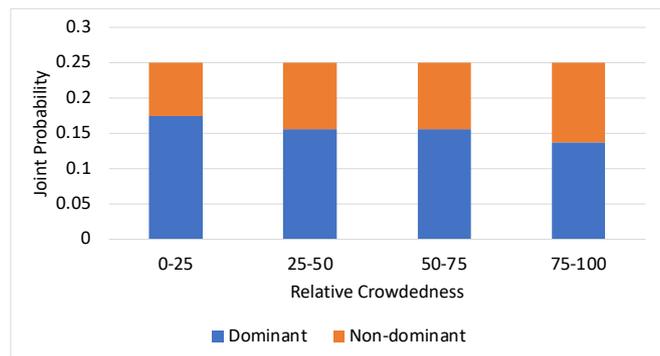

Fig.10. Joint probability distribution of assignment of preferred distance based on the crowdedness of the destination.

The assigned values of price and expected time were generated with the strong correlation. We defined four types of context consumers by price assignment: (i) defined based on the expected time to stay at the destination (i.e., Cat-1), (ii) statically defined by own preference, (iii) defined based in destination (i.e., locality), and (iv) defined based on time of the day. The proportion of context queries assigned with price and/or expected time are configurable. The Google Places API response provides the minimum (*min_time*) and the maximum (*max_time*) average number of minutes a person stays at a location. We use that data to estimate the expected time to stay (and in extension, the expected time to park) for Cat-1 context consumers. The expected time at a place follows a normal distribution, produced by the Expected Time Distribution Generator, where the average = (*min_time*+*max_time*)/2. 16% of the context query templates contains the price preference set based on the expected time at the destination. 64% of the assigned preferred price is based on a random normal distribution generated based on the price of on-street parking in Melbourne.

Finally, the developed context query templates through the multi-step process were persisted in the Context Query Template Repository (CQTR). We used this approach for the following reasons: (i) reusability of the generated context query load; apart from executing the context queries using the CQE, we also use our dataset to setup other profiles in JMeter in CSV form, (ii) provide the ability to re-run the same profile repeatedly for validation under different conditions, e.g., we use our simulator to test the performance of the CoaaS with and without caching, (iii) provide a simple interface to the CQG, and (iv) minimize the computing cost of running a large set of queries over an extended period of time in real-time (i.e., a week).

Summarizing the process above, the CQG matches a sizable number of context consumer profiles, with popularity (crowdedness) profiles of locations and the traffic volumes at different times of the day. Therefore, Fig.11 illustrates the indicators used in the context query generation process to produce a context query template set that once executed using the CQE conforms to near real-world like load patterns.

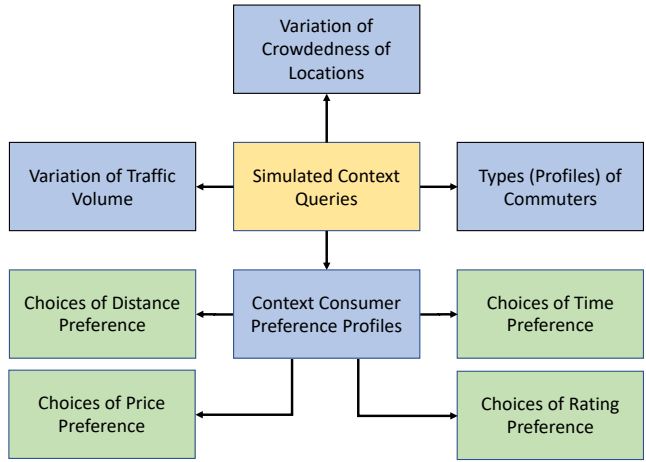

Fig.11. All the indicators used to generate the context query templates that resemble the real-world situation in MEL CBD.

One of the advantages of our context query generator for the smart car parking scenario is the query language independence. Our context query templates dataset can be used to execute context queries using any context query language. The CQG however, executes the context query load defined in the Context Definition and Query Language (CDQL) [33]. The properties defined in the dataset are shown in TABLE V.

TABLE V. Properties of the Context Query Template Dataset

| Property | Nullable | Description |
| --- | --- | --- |
| _id | False | Context query template identifier. |
| location.lat | False | Latitude of the context consumer at the time of query execution. |
| location.lng | False | Longitude of the context consumer at the time of query execution. |
| address | False | Name or address of the intended destination. |
| day | False | The specific day of the week, hour, minute, and second the context query need get executed. |
| hour | False | |
| minute | False | |
| second | False | |
| distance | True | The maximum distance the CC intends to travel from the car park to the destination. |
| expected_time | True | The expected time the CC intend to stay at the destination or park the vehicle. |
| price | True | The expected price the CC intend to pay for the duration of parking. |
| rating | True | The minimum average rating expected for car parks in the response. |
| vin | False | Vehicle identification number. |
| query_id | False | Reference to the context query in the database for the use of the CQE. |

---
[5] Available at: https://github.com/ShakthiYasas/context-query-simulator.git

We present all our generated datasets: places, vehicles, and context query templates in our repository[5].

*C. Context Query Execution*

The CQE autonomously executes the generated queries. Given that we developed our datasets for a week based on the popularity profiles of the places, the CQE is capable of executing a sizable context load through all seven days (i.e., one hundred and sixty-eight hours). The request rate ($\lambda$) of context queries follows a Poisson distribution as assigned in the generation phase.

We the CQE as a scheduler service using Quartz. The main thread executes a scheduler that gets triggered each $m$ time unit. In our case, $m = 10$ minutes. As indicated in Fig.12, the QueryFetchJob retrieves all the context query templates from the CQTR that need to be executed during the subsequent $m$ time. Then, each of these context queries are assigned to the secondary scheduler where the triggers are assigned execution time as attached in the query template. For example, when the main thread executes QueryFetchJob at 10:00am, all queries that are attached to be executed between 10:00-10:10am are retrieved. Considering query$_1$ is attached to execute at 20 seconds past 10:08 am, the process creates a one-time trigger that gets executed at this time. The complete context query string is developed at this point by merging the values from the query templates with the pre-defined query using the *query_id*. This allows us to execute different context queries. Fig.13 depict a different context query that context-aware search for available parking slots, different to that previously shown in Fig.1. For simplicity however, we used only a single query, i.e., indicated in Fig.1 with optional conditions, when executing our simulator. We implement the publisher-subscriber pattern to listen to trigger executions, and hence the listening QueryJob makes a HTTP POST request to CoaaS.

## V. CONCLUSION

Deriving reliable context information from IoT data is increasingly possible by the day with a large number of devices being deployed. Possible use cases utilizing context information are endless, enabling the development of smarter applications. There is a compelling need to investigate scalable techniques for Context Management Platforms to infer and deliver context to a large number of heterogenous context consumers as a result. However, there is limited evidence of investigations in scalability of CMPs and/or experimentation conducted using a sizable context query load. In this paper, we propose a context query simulation process considering smart car parking in the Melbourne Central Business District as a use case. The simulation process consisting of two services, context query generation and execution, is context query language independent, and produces context queries that resemble a real-world scenario. The simulated context query load conforms to patterns and features derived using real traffic variations, crowdedness of places and different profiles of commuters. We present our algorithms and a dataset of 898,050 context query templates which could be used to simulate a context query load over a period of seven days. Hence, our dataset is significantly large enough, in order to provide significant volumes of experimental data to reliably test and produce credible result about scalability functions, serving as a potential benchmark for context query processing research. We find adaptive context caching, auto scaling, and distributed context management using edge devices as areas of research that our simulator would be useful for testing. Further areas of research that benefit from the simulator are investigation into approaches to identifying patterns in context information, e.g., identifying follow up queries after querying for parking slots for proactive context caching, and detecting features of context query loads for personalization, e.g., features of recurrence context queries searching for parking slots during the rush hour.

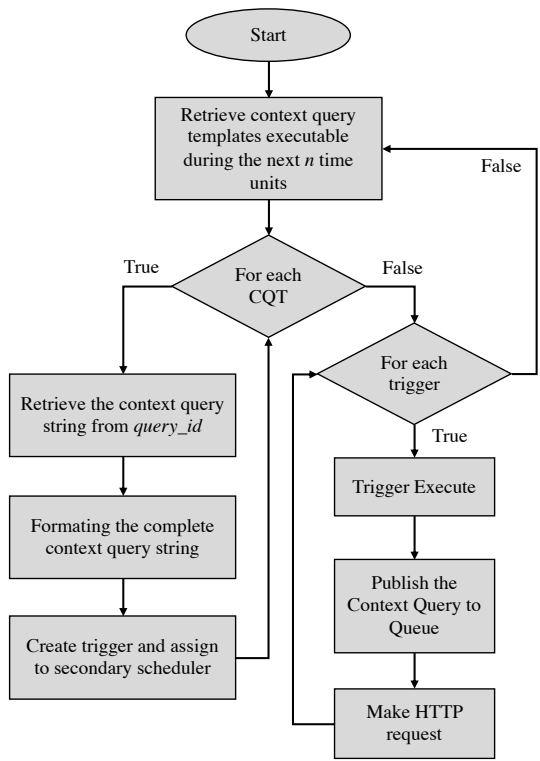

Fig.12. Process of the Context Query Executor (CQE).

```
prefix schema:http//schema.org
push (targetCarpark.*)
when
  distance(consumerCar.location,
targetLocation.geo)<={"value":500, "unit":"m"}
define
entity targetLocation is from schema:Place
  where targetLocation.name="Melbourne Skydeck",
entity consumerCar is from schema:Vehicle
  where cosumerCar.vin="13UNVER82367G4",
entity targetCarpark is from schema:ParkingFacility
  where
    goodForWalking(targetWeather)>=0.6 and
    targetCarpark.maxHeight>consumerCar.height
    targetCarpark.isOpen=true and
    targetCarpark.availableSlots>0
```

Fig.13. An alternate context query to search available car parking slots defined in CDQL [33].

ACKNOWLEDGMENT

Support for this publication from the Australian Research Council (ARC) Discovery Project Grant DP200102299 is thankfully acknowledged.